\documentclass[runningheads]{llncs}
\usepackage{llncsdoc}
\usepackage{subfig}
\usepackage{graphicx} % for pdf, bitmapped graphics files

\usepackage{epsfig} % for postscript graphics files
\usepackage{amsfonts}
\usepackage{amsmath}
\usepackage{amssymb}

\begin{document}

\title{Haptic Rendering of Thin, Deformable Objects with Spatially Varying Stiffness}
\titlerunning{Haptic Rendering under Stiffness Variation}

\author{Priyadarshini Kumari \and Subhasis Chaudhuri}
%\authorrunning{I. Ekeland,  R. Temam}
\institute{Vision and Image Processing Laboratory, Department of Electrical Engineering. Indian Institute of Technology Bombay, Powai, Mumbai-400076\\
\email{\{priydarshini,sc\}@ee.iitb.ac.in}}
%\and
%Department2, University2, address, Country2\\
%\email{roger@smwhere.com}}

\maketitle

%%%%%%%%%%%%%%%%%%%%%%%%%%%%%%%%%%%%%%
\begin{abstract}
In the real world, we often come across soft objects having spatially varying stiffness, such as human palm or a wart on the skin. In this paper, we propose a novel approach to render thin, deformable objects having spatially varying stiffness (inhomogeneous material). We use the classical Kirchhoff
thin plate theory to compute the deformation. In general,
the physics-based rendering of an arbitrary 3D surface is complex
and time-consuming. Therefore, we approximate the 3D surface
locally by a 2D plane using an area-preserving mapping technique - Gall-Peters mapping. Once the deformation is computed
by solving a fourth-order partial differential equation, we
project the points back onto the original object for proper
haptic rendering. The method was validated through user
experiments and was found to be realistic.

\end{abstract}

%%%%%%%%%%%%%%%%%%%%%%%%%%%%%%%%%%%%%%
\section{Introduction}
Haptic perception becomes more realistic when we incorporate the physical and material properties of an object. For the kinesthetic rendering of rigid objects, it is sufficient to calculate only the force as they do not undergo any deformation under the applied force. However, the rendering of deformable objects is more complex since we need to calculate the force as well as the deformation. If the stiffness of the object varies over the surface, there is further complications as the rendered force is a function of both local deformation and local variation in stiffness. The continuity of deformation and a smooth movement of the proxy on the deforming surface are some of the key concerns while rendering a deformable object having a variable stiffness.\par
In the last two decades, haptic rendering has found several applications on interactions with the virtual environment, such as virtual museum\cite{cultural}, dental surgery simulation\cite{dentalsurgery}, and force modeling for needle insertion\cite{needleinsertion}. However, these interactions cannot handle skin or bowel simulator, which requires the object to be both deformable and have variable stiffness. The proposed method tries to achieve this special type of rendering.\par 
We propose a method to render a point cloud data representing thin deformable objects having spatially varying stiffness such as skin and elastic sheets. First, the point cloud in the neighborhood of the proxy is locally fit into a hemispherical surface and using the parameter of the fitted sphere these points are then projected on a plane using Gall-Peters projection\cite{peter}. This being an area-preserving mapping, preserves the stiffness variation on the projected 2D plane.  Subsequently, we use the Kirchhoff's thin plate theory\cite{thinplate} to calculate the deformation due to an applied force at the proxy and for a given stiffness pattern. The solution is obtained by solving a fourth-order non-homogeneous Partial Differential Equation (PDE). Once the deformation map is estimated, it is back projected on the original 3D surface for proper rendering. \par
Most of the work done in the field of deformable object rendering can be classified into two categories: geometry-based deformation and physics-based deformation. Geometry based models are fast but do not necessarily focus on the physics involved in the deformation. On the other hand, physics-based rendering model is computationally expensive as the simulation of physical and material properties of the object is complex.
In geometry-based modeling most of the methods use Gaussian\cite{gaussian} or a polynomial function to move the surrounding vertices around Haptic Interaction Point (HIP). Another widely practiced method is spline-based in which control points are assigned on the object, and these control points are manipulated to effect a smooth deformation \cite{geometry1}, \cite{geometry2}.\par

A FEM based method, such as \cite{SalisburyMedical} provides the most accurate rendering. However, this is computationally very demanding for haptic applications, requiring specialized hardware. It has been shown in \cite{FEM_MS} that by sacrificing the accuracy marginally, the computations can be speeded up by a factor of 10 by using a mass-spring model. However, the choice of dampers is very critical to ascertain the desired dynamic behavior, and usually, they are chosen heuristically.  The purpose of the proposed method is to avoid these heuristics by solving a proper PDE over a 2D plate.

In smoothed-particle-hydrodynamics (SPH) approach\cite{SPH}, authors have introduced a technique that renders the object in different possible states (fluid, elastic, and rigid) in the same scene. However, in this technique, the stiffness of the entire object is varied as opposed to spatially varying the stiffness.
Moreover in \cite{SPH}, a volume model of the object is used, while we use a 3D surface model which is comparatively much faster than the volume model.\par
The key contribution of our work is the development of an alternate, stable, physics-based haptic rendering method for an inhomogeneous, elastic deformable object.

\section{Proposed Method}
\begin{figure*}[thpb]
      \centering
        \includegraphics[scale=0.65]{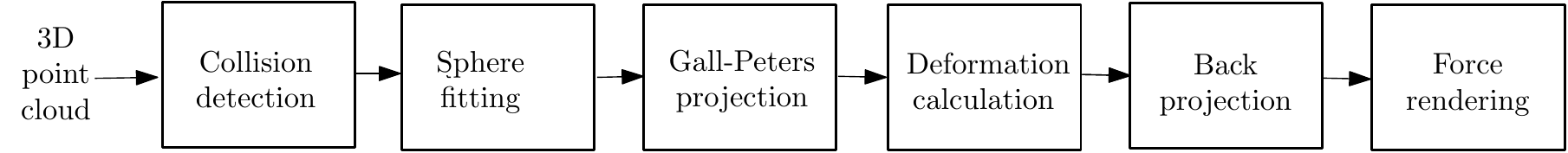}
      \caption{Various steps involved in haptic rendering of a deformable object with variable stiffness.}
      \label{block}
    \end{figure*}
The input for our method is the point cloud representation of the 3D object. Since there is a one-to-one relationship between the point cloud and mesh data\cite{pointmesh} (albeit the rendering process is a bit different), the algorithm can be extended to other forms of data representation. The point cloud is given by  4-tuple point $ {P}_i =\left \{({x}_i, {y}_i, {z}_i), {d}_i \right \}$, where $({x}_i, {y}_i, {z}_i)$ is the location of the $i^{\text {th}}$ point and ${d}_i$ is the corresponding stiffness. The overall algorithm can be broken into several stages as shown in Fig. \ref{block}.
The various stages of algorithm are described in the following subsections.

\subsection{Collision Detection} \label{Collision detection}
\begin{figure}[thpb]
   \centering
   \includegraphics[scale=0.5]{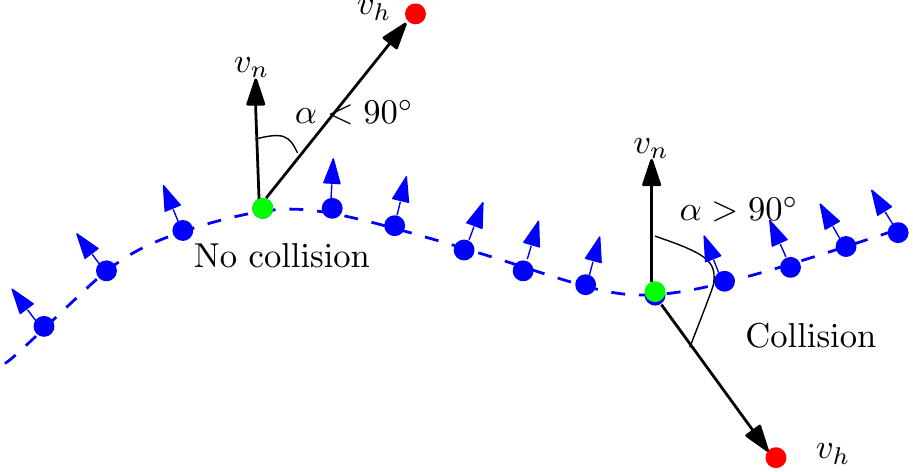}
   \caption{Illustration of collision detection. The proxy and the HIP are shown in green and red colors, respectively. The point cloud is shown in blue. Blue arrows show the computed outward surface normals.}
   \label{collision}
\end{figure}

With point cloud data, we do not have surface normal defined at a point. In order to provide the appropriate force feedback to the user, we need the normal information at the collision point. We follow a proxy-based technique proposed in \cite{normal} for computation of normals. In this method, the normal is computed from the radial overshoot of each point inside the spherical proxy. Once the collision occurs, HIP penetrates the object, and the proxy is constrained to lie on the surface by using a dynamic function as defined in \cite{normal}. A collision is detected when the condition $(\mathbf{v}_n.\mathbf{v}_h)<0$ is satisfied, where $\mathbf{v}_n$ is the computed surface normal at the point of collision and $\mathbf{v}_h$ is the vector joining the proxy to HIP as shown in Fig. \ref{collision}. 
Once the collision point is detected, we select the point cloud in the neighborhood of the point of collision for fitting a sphere locally.
\subsection{Spherical Facet Modeling}

We use the algebraic distance-based method proposed in \cite{pratt} to fit the surface points in the neighborhood of the proxy. Let ${f({x}, {y}, {z})}$ be an equation of sphere in the algebraic form

\begin{equation} \label{algebraicSphere}
\begin{split}
{f({x}, {y}, {z})} =  {u}_0({x}^{2}+{y}^{2}+{z}^{2})+{u}_1{x}+{u}_2{y}+{u}_3{z}+{u}_4=0 
\\ 
\text{subject to} \hspace{0.4cm} {u}_1^2+{u}_2^2+{u}_3^2-4{u}_0{u}_4 = 1\hspace{0.9cm}
\end{split}
\end{equation} 
where ${\mathbf{u}}$ = $[{u}_0 \hspace{0.2cm} {u}_1 \hspace{0.2cm} {u}_2 \hspace{0.2cm} {u}_3 \hspace{0.2cm} {u}_4]^{T}$ represents the coefficients describing the  sphere. The above mentioned constraint can be written in the matrix form as ${\mathbf{u}}^T\mathbf{C}{\mathbf{u}} = 1$  with $\mathbf{C}$ as given in (\ref{unconstraint}). So any point $({x}_i, {y}_i, {z}_i)$ lying on the surface of sphere will have an algebraic distance equal to zero. We minimize the sum of squared algebraic distances to estimate the best possible spherical fit ${\mathbf{u}}$. Although several possible constraints have been suggested in \cite{bookstein},\cite{albano}, we use the constraint ${\mathbf{u}}^T\mathbf{C}{\mathbf{u}} = 1$, as Pratt \cite{pratt} has shown that such a constraint has only one point of singularity of being a zero radius sphere. The resulting cost function can be expressed by introducing a Lagrangian multiplier $\lambda$ as:
\begin{equation} \label{unconstraint}
\underset{\mathbf{u}}{\text{minimize}}\hspace{0.1cm} ( \|\mathbf{D}{\mathbf{u}}\|^2 - \lambda ({\mathbf{u}}^{T}\mathbf{C}{\mathbf{u}} - 1))
\end{equation}

\[\mathbf{D} = 
\begin{bmatrix}
{x}_1^{2}+{y}_1^{2}+{z}_1^{2} & {x}_1 & {y}_1 & {z}_1 & 1 \\
\vdots & \vdots & \vdots & \vdots & \vdots \\
{x}_n^{2}+{y}_n^{2}+{z}_n^{2} & {x}_n & {y}_n & {z}_n  & 1\\
\end{bmatrix}
,
\mathbf{C} = 
\begin{bmatrix}
0 & 0 & 0 & 0 & -2 \\
0 & 1 & 0 & 0 & 0 \\
0 & 0 & 1 & 0 & 0 \\
0 & 0 & 0 & 1 & 0 \\
-2 & 0 & 0 & 0 & 0 \\
\end{bmatrix}
\]
where $\mathbf{D}$ is an $n\times5$ matrix derived from $n$ points from the point cloud data, selected in the neighborhood of the collision point. 

The solution is given by solving the generalized eigenvalue problem $\mathbf{D}^T \mathbf{D}\mathbf{u}$ = $ \lambda \mathbf{C}{\mathbf{u}}$, which can be solved using QZ algorithm developed by Moler and Stewart \cite{QZ}.
The center and the radius of the sphere can be calculated from the components of ${\mathbf{u}}$ as given in \cite{algebraic}

% center of sphere
\begin{equation} \label{center}
{\mathbf{c}}= {-\frac{ 1 }{ 2{u}_4 }} [{u}_1 \hspace{0.3cm} {u}_2 \hspace{0.3cm} {u}_3],\hspace{0.1cm} {r}=\sqrt{{c}^T{c}- {\frac{{u}_0}{{u}_4}}}.
\end{equation}
Once we get the center $\mathbf{c}$ and the radius $r$ of the sphere, all the points are locally projected on to a 2D plane as discussed in the next subsection.\par
 
\begin{figure}[thpb]
  \centering
\subfloat[][]{\includegraphics[scale=0.35]{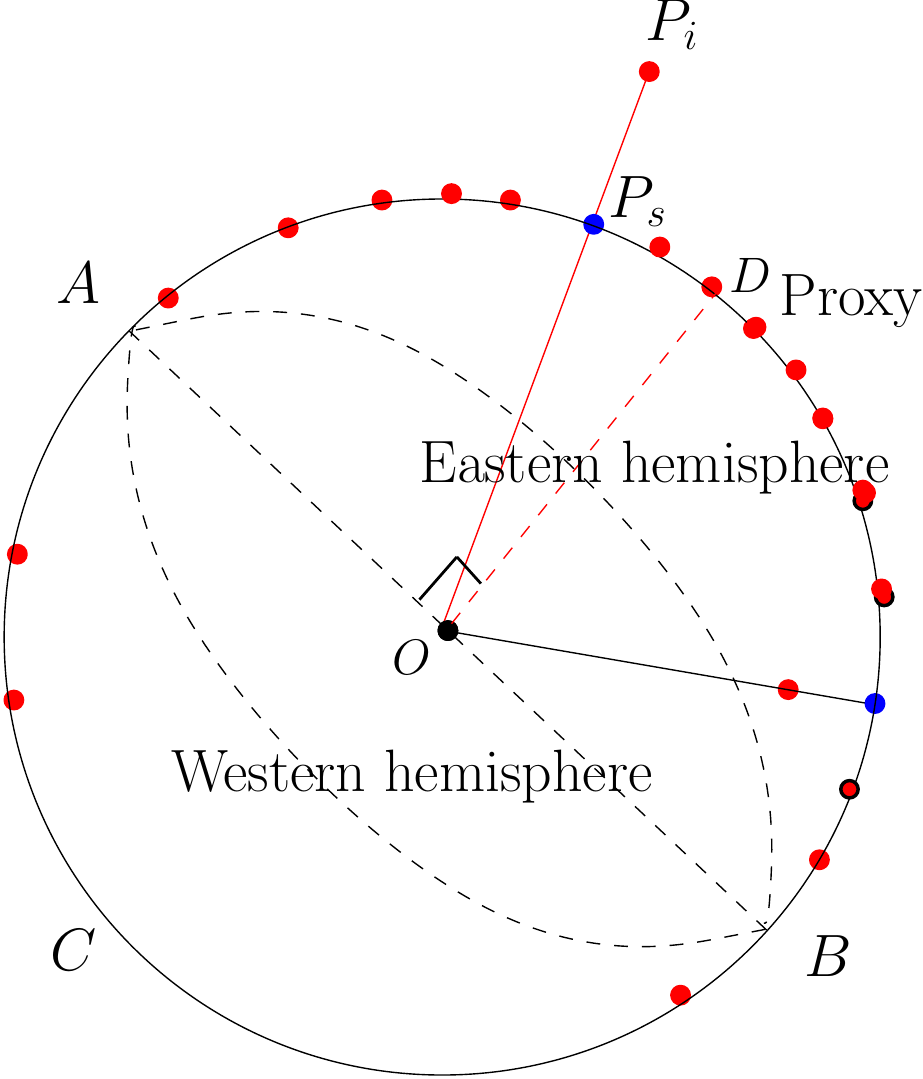}\label{projectedPoint}} \hspace{0.8cm}
\subfloat[][]{\includegraphics[scale=0.3]{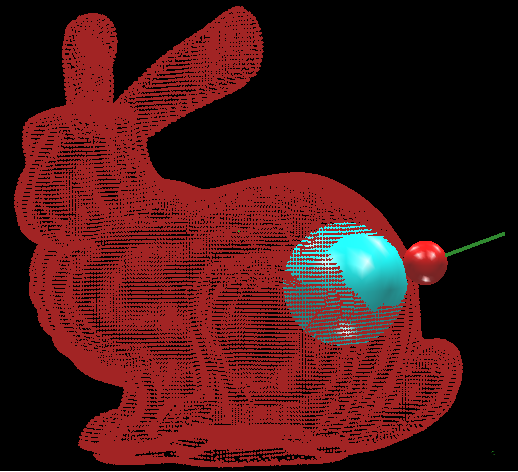} \label{SphereFit}}
\caption{(a) Illustration of projection of points on the surface of the best fit sphere and employing the constraint of hemisphere fitting. The object points ${P}_i$ are shown in red, and the projected points ${P}_s$ are shown in blue. (b)Illustration of the best fit sphere to the set of object points near the proxy. The proxy is shown with a red ball, and the green line points to the direction of normal at the collision point. (Data Courtesy: http://graphics.stanford.edu/data/3Dscanrep/)}
 \label{latbound}
\end{figure}

All points, however, need not lie on the surface of the best fit sphere, as shown in Fig. \ref{projectedPoint}. In order to approximate the object surface by an area-preserving planar patch, we need to project all points on the surface of the fitted sphere. We use a projection technique in which we join all the points (which are not lying on the surface of the sphere) from the center. The joining line intersects the sphere at two points. The intersecting point nearest to the object point is chosen to be the projection of the corresponding point on the sphere.  Figure \ref{SphereFit} shows the best fit sphere to a set of object points near the proxy.

As it was mentioned earlier that deformation due to an arbitrary variation in material properties is easier to compute for a thin sheet, the fitted sphere is mapped onto a 2D plane. For an equi-curvature surface (like a sphere), the amount of energy required for deformation is proportional to the surface area when the stiffness is constant. Hence we require a projection operator $g:{\rm I\!R}^3\rightarrow{\rm I\!R}^2$, which is area-preserving so that deformation computation is a function of stiffness alone. Gall-Peters projection \cite{peter} used in cartography is one such operator that preserves the area. Hence we use Gall-Peters projection in a way so that the proxy position lies at the center of the unfolded globe.% (see Fig. \ref{boundaryC1}).
 While fitting the sphere, if there are points on the western hemisphere, these points are not visible at the proxy, and hence they are rejected, and only a hemispherical facet is considered. This is illustrated in Fig. \ref{projectedPoint}, where $D$ is the proxy on the surface, $AB$ corresponds to the plane joining the north pole ($A$) and the south pole ($B$) and the point $D$ corresponds to $(0, 0)$ latitude and longitude of the sphere. Points lying on the western hemisphere (shown as the arc $ACB$) are discarded. 

\subsection{Area Preserving 3D to 2D Projection}

Since Gall-Peters projection describes a technique to project the globe onto a plane, we have used it for a sphere to plane projection. The longitude and latitude of each point on the sphere is calculated from polar coordinates of the points. Considering the z-axis to be perpendicular to the computer screen, each point on the sphere is projected onto the x-z plane using  (\ref{mapProjection}).
%equaion of map projection
\begin{equation} \label{mapProjection}
%\begin{split}
x=\frac{r\alpha}{\sqrt{2}}\\ \hspace{1cm} z=r\sqrt{2}sin(\beta),
%\end{split}
\end{equation}
where $\alpha$ and $\beta$ are longitude and latitude of the point in radians respectively and $r$ is  radius of the sphere.
While using the Gall-Peters projection some issues emerge, when points lie either near the  pole region or near the 180$^{\circ}$ longitude. 
\begin{enumerate}
\item A point lying on the north/south pole gets stretched to a line on the plane after projection. 
\item Points on 180$^{\circ}$ longitude get mapped on both the extreme of the 2D plane.
\end{enumerate}
We avert such problem by shifting the coordinate axis of sphere so that our proxy (point $D$ in Fig. \ref{projectedPoint}) falls in the region of zero latitude and zero longitude. After shifting the axes, we discard the points on the other half of the sphere opposite to the proxy (arc $ACB$ in Fig. \ref{projectedPoint}).  
In this way, all neighborhood points around the proxy fall near the center of the plane, which would be very useful while employing an infinite horizon boundary condition in section \ref{Boundaries conditions for PDE}.
Once the points are projected on a plane, we compute the deformation of each point using Kirchhoff thin plate theory \cite{thinplate}. As mentioned earlier, computation of deformation on the planar sheet is easier than directly computing on the object surface. Therefore we compute the deformation of each point on 2D plane and subsequently project the points back on the original 3D object. 

\subsection{Deformation Computation on a  Planar Sheet}

In Kirchhoff theory of plate \cite{thinplate}, the thickness of the plate is assumed to be very small as compared to the planar dimensions. In addition, some more assumptions are made to reduce the three dimensional problem to two dimensions. These assumptions are summarized as follow 
\begin{enumerate}
\item The line normal to the neutral axis remain straight after deformation.
\item The normal stress in the direction of thickness is neglected.
\item Thickness of the plate does not change during bending.
 \end{enumerate}
Let ${s}$, ${w}$ and ${t}$ be the displacements of a point along x, y and z directions, respectively, and $\left | \mathbf{F} \right |$ is a distributed load on the planar sheet. 
Since the planar sheet is on the x-z plane, we can write the variation of ${s}$ and ${t}$  across thickness in terms of displacement ${w}$ as $s=-y\frac{\partial w}{\partial x}, \; t=-y\frac{\partial w}{\partial z}$.

Taking all the assumptions into account, the deformation of a point is governed by the following PDE as explained in \cite{thinplate}.
\begin{equation} \label{differentialC}
%\begin{split}
\frac{\partial^4 w}{\partial x^4} +2 \frac{\partial^4 w}{\partial x^2\partial z^2} + \frac{\partial^4 w}{\partial z^4}=\frac{\left | \mathbf{F} \right |}{D},\hspace{0.3cm}
or\hspace{0.3cm} {\nabla^4 w} = \frac{\left | \mathbf{F} \right |}{D}, \hspace{2cm}
%\end{split}
\end{equation}
where $\nabla^4$ is the biharmonic operator and $D$ is flexural rigidity of the object. Interestingly, this equation was first obtained by Lagrange in 1811. The flexural rigidity is a material property of the object and is defined as
\begin{equation} \label{flexural}
D=\frac{Eh^3}{12(1-v^2)},
\end{equation}
where $E$, $h$, $v$ are the modulus of elasticity, thickness of the plate and Poisson's ratio (usually $0<v<0.3$), respectively. Equation (\ref{differentialC}) is valid only if material property $D$ of the object is constant. In this case, computation of deformation also becomes easier as the governing differential equation (\ref{differentialC}) is a standard biharmonic equation.  However in real world, there are numerous objects which have spatially varying material properties when $E$ and $v$ are functions of $x$ and $z$. Equation (\ref{differentialC}) cannot be applied to compute deformation for such objects. Hence we move to the next stage where stiffness of the object is allowed to vary over the surface. \par
In order to compute the deformation in variable stiffness object we use the extended Kirchhoff thin plate theory\cite{thinplate}. 
The deformation of each point in such cases is governed by the following equation
\begin{equation} \label{differential2}
\begin{split}
D{\nabla}^4w + 2\frac{\partial D}{\partial x}\frac{\partial }{\partial x}({\nabla}^2w) + 2\frac{\partial D}{\partial z}\frac{\partial }{\partial z}({\nabla}^2w) + {\nabla}^2D({\nabla}^2w) \\
-(1-v)\left(\frac{\partial^2 D}{\partial x^2}\frac{\partial^2 w}{\partial z^2} -2\frac{\partial^2 D}{\partial x\partial z}\frac{\partial^2 w}{\partial x\partial z} + \frac{\partial^2 D}{\partial z^2}\frac{\partial^2 w}{\partial x^2}\right) = {\left | \mathbf{F} \right |}.
\end{split}
\end{equation}
The detail derivation of  (\ref{differentialC}) and (\ref{differential2}) is explained in \cite{thinplate}.
Unfortunately finding an analytical solution of this form of PDE is extremely difficult. Hence we solve the equation in discrete domain by using Jacobi iterative method for a given boundary condition. 
As the Gall-Peters projection uses non-linear mapping function (\ref{mapProjection}),  points are not projected evenly on a plane. Therefore we first sample the projected points on the uniform grid and 
subsequently we discretize all the variables and parameters of  (\ref{differential2}) by using central difference. The value of $D$ at a grid node is taken as the value of $D$ of the closest projected point from the grid node. 
Our initial condition on deformation is $w(i,j)=0$. The final expression for iterative updating can be written in the form of a function $L()$:
\begin{equation} \label{discrete2}
w^{n+1}(i,j)=L(\left | {F} \right |, \mathbf{D}(i,j), w_{\mathcal{N}}^{n}(i,j))
%{w(i,j)}^{n+1}=f(p, D, ...w(i-1,j+1)^{n}, w(i,j-1)^{n}, \\w(i,j+1)^{n}, w(i+1,j-1)^{n},...)
\end{equation}
where the superscript $n$ denotes the $n^{th}$ iteration and $w_{\mathcal{N}}^{n}(i,j)$ refers to the various lattice entries in the neighborhood of the central lattice $(i,j)$,  while $\mathbf{D}(i,j)$ is the given flexural rigidity map and its all derivatives (upto second order) at the lattice $(i,j)$. In each iteration, we update the $w$ matrix and compare it with  $w$ matrix of previous iteration.
We stop the iteration when the Frobenius norm of difference between current and previous  $w$ matrices becomes very small.
The final $w$ matrix gives the deformation at the grid point. However, one needs the deformation value at the location where the 3D point was originally projected using  (\ref{mapProjection}).
The deformations at projected points are computed using bi-linear interpolation.

%Boundaries conditions for PDE
\subsection{Boundary Conditions for PDE} \label{Boundaries conditions for PDE}
As we mentioned earlier, in order to obtain the solution of the PDE given in  (\ref{differential2}), we need suitable boundary conditions. For simplicity, we assume edges of the projected plane to be parallel to the coordinate axis $X$ and $Z$.

We need two boundary conditions at each edge. In our work, we assume all four edges of projected plate to be fixed (i.e deformation at infinite horizon is zero). Hence the deflection and the first order derivative become zero at all four edges. After getting the deformation locally on the projected plane, we project back the points to compute the deformation on the original object for proper rendering. Points on the object are displaced in the direction of the force $\mathbf{F}$ using the governing equation (\ref{backProj})
\begin{equation} \label{backProj}
\end{equation}
where $\hat{P}_i$ is the final position of the $i^{\text{th}}$ point on the deformed surface and $P_i$ corresponds to the position before deformation with $w_{i}$ being the deformation at the point. 

%Force rendering  
\section{Force Rendering} \label{Force rendering}
In order to haptically render the object, we first detect the collision of HIP with the object as discussed in section \ref{Collision detection} and compute the force if a collision is detected. 
Once the collision is detected, the force needs to be fed back by the haptic device to provide the sensation of touch. Two important factors for force computation are its magnitude and its direction. The magnitude of force should be proportional to the penetration depth of HIP from the surface and its direction should be in the direction of the normal at the point of collision. Hence the reaction force is calculated using the expression, $F =-\frac{EA}{h}(\left|{X}_h-{X}_p\right|)$. Here ${X}_h$ is the HIP position, ${X}_p$ is the proxy position, $A$ is the area of the plate and $E$ is the modulus of elasticity at that location.

%Results
\section{Results}
 \begin{figure}[t!]
   \centering
\subfloat[][]{\includegraphics[scale=0.25]{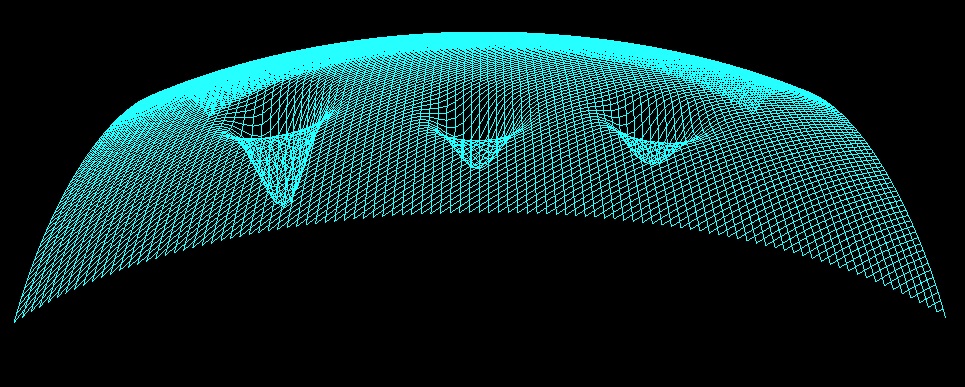}\label{linear}}\hspace{0.3cm}
\subfloat[][]{\includegraphics[scale=0.25]{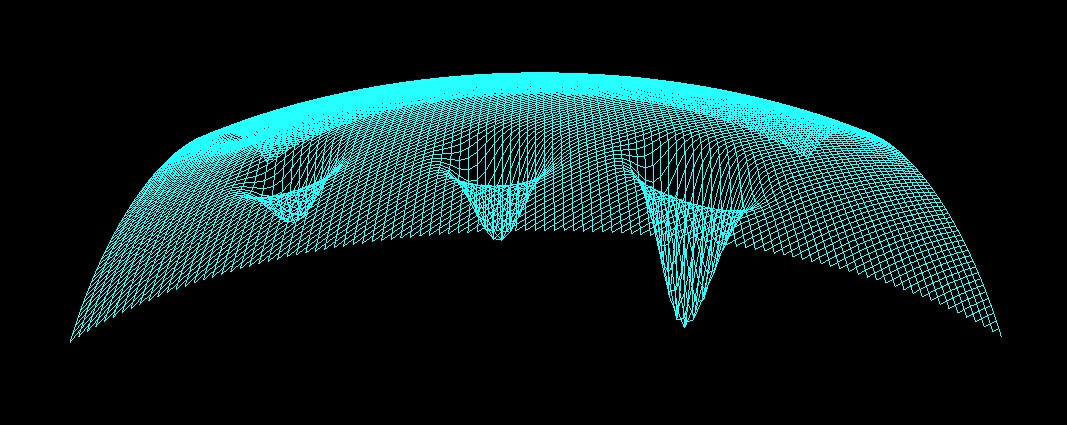}\label{hyper}}
\caption{Computed deformations on the object surface with different flexural rigidity functions under the same amount of applied force: (a) Linear, (b) Hyperbolic ($x$ in cm). Deformation at various locations are superimposed together for ease of visualization.}
 \label{def}
\end{figure}
    
The proposed method was implemented in Visual Studio 2010 in a Windows 7 platform with a Core(TM2) Quad CPU Q8400 processor @ 2.66 GHz clock speed with 8 GB RAM. We use OpenGL 2.0 for graphic rendering and HAPI library for haptic rendering. We experimented with 3D point cloud model using Falcon haptic device from NOVINT. In order to make the rendering faster we segregate the tasks of deformation computation, proxy updation and force computation through three different threads while programming. We observed that the deformation thread, as expected, runs relatively slowly due to iterative solution for equation (\ref{differential2}). The average time required to run the deformation thread once is around 160 ms. We have set $v=0.2$ in the study.
\begin{figure}[t!]
   \centering
   \includegraphics[scale=0.25]{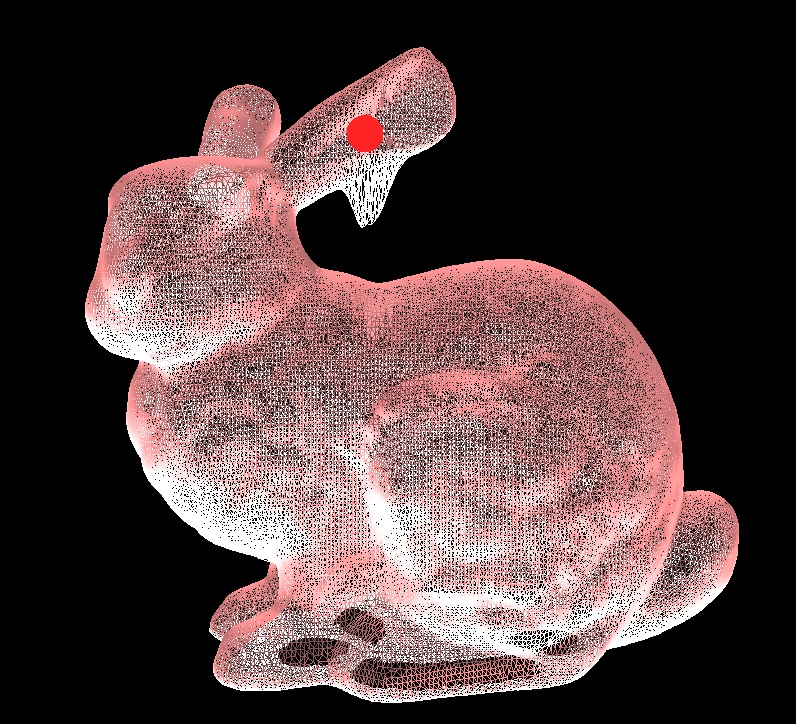}
   \caption{Deformation shown on the bunny model.}
   \label{bunny}
\end{figure}
%\subsection{Visualization}
\par
We initially use a simulated elliptical object consisting of 40000 points. The 3D grid used to sample the point cloud is of size $200\times200\times200$ with inter-node space of 0.025cm.
Figure \ref{def} shows the deformations at several points on the object surface with different flexural rigidity functions. In Fig. \ref{linear}, flexural rigidity is varied gradually along the horizontal direction while in Fig. \ref{hyper} it is varied rapidly. We applied the  same amount of force on the object surface and observed that deformation decreases as flexural rigidity increases as shown in Fig. \ref{linear}. It may be noted that the deformation is restricted to the neighborhood only due to the choice of boundary conditions. To explain this further, we experiment on a real object model and show the deformation at the ear lobe of a bunny in Fig. \ref{bunny}. One may expect the ear lobe to be bent due to the applied force like a cantilever. Instead one observes a dip within the ear lobe, exact shape of which would depend on the size of the chosen neighborhood.\par

We also verify the performance of the proposed algorithm in terms of user experience while interacting with various object models. We set up an experiment in which users were asked to interact with an object model (say Fig. \ref{bunny}) and rate the experience on a scale of $1-5$, with 1 being very poor to 5 being excellent.  Most of the users (80\%) rated the experience as very good (rating 4) remaining 20\% gave the rating of 3. When asked if they are able to perceive the change in stiffness, 90\% users responded positively.

\section{Conclusion}
We have proposed a numerical method to solve the haptic rendering problem of elastic  deformable objects with spatially varying stiffness. It is a physics-based rendering technique and gives a realistic feeling of touching the object. We have not observed any divergence issue in the computation of the deformation map or the proxy movement over the deforming surface. In order to mathematically handle any object, an area-preserving map projection technique is used to transform an arbitrary 3D surface to a planar one following which Kirchhoff's thin plate deformation propagation model involving a fourth-order PDE is used to compute deformation in case of spatially varying stiffness. We also tested with some subjects and observed that the hapto-visual rendering technique using the proposed algorithm greatly augmented the user's experience. Our future work will involve the choice of alternate boundary conditions and 
extension to dealing with volumetric data. 

\bibliographystyle{splncs03}
\bibliography{bibliography_file}

\end{document}